\documentclass[reprint,onecolumn,longbibliography,eqsecnum,floats,showpacs,%
nofootinbib,amsmath,amssymb,aps,prd,superscriptaddress,notitlepage]{revtex4-1}

\usepackage{graphicx}% Include figure files
\usepackage{dcolumn}% Align table columns on decimal point
\usepackage{bm}% bold math
\usepackage{hyperref}% makes the references hyperlinked

\usepackage[T1]{fontenc}

\newcommand{\re}{\mathbb{R}}
\newcommand{\ig}{\mathbb{Z}}
\newcommand{\rd}{{\rm d}}
\newcommand{\rx}{{\rm x}}

\newcommand{\POrd}{\mathcal{P}}

\newcommand{\Hil}{\mathcal{H}}
\newcommand{\kin}{{\rm kin}}

\newcommand{\phy}{{\rm phy}}

\newcommand{\LQG}{{\rm LQG}}
\newcommand{\LQC}{{\rm LQC}}
\newcommand{\diff}{{\rm diff}}
\newcommand{\Bohr}{{\rm Bohr}}
\newcommand{\Ar}{\rm Ar}

\newcommand{\Pl}{{\rm Pl}}
\newcommand{\lPl}{\ell_{\Pl}}

\newcommand{\oq}{{{}^o\! q}}
\newcommand{\ofe}{{{}^o\! e}}

\newcommand{\ket}[1]{{| #1 \rangle}}
\newcommand{\bra}[1]{{\langle #1 |}}

\usepackage{colordvi}

\usepackage[active]{srcltx}

\begin{document}
\preprint{}

\title{Observations on interfacing loop quantum gravity with cosmology}

\author{Tomasz \surname{Paw{\l}owski}}
	\email{tomasz.pawlowski@unab.cl}
	\affiliation{Departamento de Ciencias F\'isicas, Facultad de Ciencias Exactas,
    Universidad Andres Bello, Av.~Rep\'ublica 220,  Santiago 8370134, Chile.}
  \affiliation{Instytut Fizyki Teoretycznej, Uniwersytet Warszawski, Pasteura 5, 
		02-093 Warszawa, Poland, EU.}

\begin{abstract}
	A simple idea of relating the LQG and LQC degrees of freedom is discussed in 
	context of toroidal Bianchi I model. The idea is an expansion of the construction
	originally introduced by Ashtekar and Wilson-Ewing and relies on explicit averaging
	of certain sub-class of spin-networks over the subgroup of the diffeomorphisms 
	remaining after the gauge fixing used in homogeneous LQC. It is based on the set 
	of clearly defined principles, thus is a convenient tool to control the emergence 
	and behavior of the cosmological degrees of freedom in studies of dynamics in 
	canonical LQG. Relating the proposed LQG-LQC interface with some results 
	on black hole entropy suggests a modification to the \emph{area gap} 
	value currently used in LQC.
\end{abstract}


\maketitle

\section{introduction}

Loop Quantum Gravity \cite{t-lqg,r-qg,al-lqg-rev} (LQG) --one of the leading 
attempts to provide a solid framework unifying General Relativity with quantum 
aspects of reality-- have matured over the past years to the level, in which 
extracting the concrete dynamical predictions out of it became a technically 
feasible task \cite{gt-aqg4,*dgkl-gq,hp-lqg-prl,*s-comm,*gt-matref}. Despite 
this success the concrete results regarding the evolution of quantum spacetime 
in LQG are yet to appear. On the other hand, in past ten years a series of 
dynamical predictions
have been made (with various level of rigor) within the symmetry reduced framework
originating from LQG known as Loop Quantum Cosmology \cite{b-livrev,*bcmb-rev,*as-rev} 
(LQC). There, the set of already known results ranges from establishing a singularity 
resolution \cite{b-sing} through qualitative changes of the standard early universe 
dynamics picture (found on the genuine quantum level) \cite{aps-prl,*aps-det,%
*apsv-spher,*acs-aspects,*bp-negL,*pa-posL,*v-open,*hp-lqc-let,*ppwe-radiation,%
*mbmmp-b1-evo} to predictions of the behavior of cosmological perturbations 
\cite{bhks,*aan-pert1,*aan-pert2,*aan-pert3,*cmbg,*lcb,*we-pert1,*we-pert2,%
*fmmmo-pert1,*fmmmo-pert2} and (in some cases) nonperturbative inhomogeneities 
\cite{bmp-geff-ftc,*bmp-geff-det}. 

The results obtained within LQC framework cannot be treated as final, as 
LQC was never derived from LQG in any systematic way. Instead, it is a
stand-alone theory constructed by applying the methods of LQG to cosmological 
models \cite{abl-lqc} further enhanced via parachuting some results and properties 
of LQG on the phenomenological level \cite{aps-imp}. 
Therefore, it is not a priori clear to what extent (if any) the predictions 
mentioned above reflect the true features of full LQG. Addressing this issues has 
brought a considerable interest within the loop community. Its pioneering studies 
\cite{bck-ref} were aimed towards controlling the so called \emph{inverse volume 
corrections} in LQC and as tools to control the heuristic effective descriptions 
of inhomogeneous extensions of LQC \cite{bhkss-str}.
Presently, the attempts to provide 
a precise connection between LQG and LQC are directed in three main areas:
$(i)$ direct embedding of LQC framework within LQG one, 
$(ii)$ approximation of cosmological (symmetric) solutions in LQG, and 
$(iii)$ emergence of cosmological (LQC) degrees of freedom in appropriate scenarios 
within LQG. 

The approach $(i)$, realized on the level of mathematical formalism, focuses on 
embedding of the elements of the LQC formalism (for example straight holonomies) 
as a proper subclass of their analogs in LQG. So far however the most natural ways
of constructing such embeddings have proved to lead to inconsistencies and resulted 
in several \emph{no-go} statement \cite{bf-config,*f-symm,*bf-nonembed}. The main 
problem encountered in the attempts is the inherent diffeomorphism-invariance of
LQG \cite{almmt-diffinv} and the fact that the symmetries characterizing the cosmological 
solutions are a subgroup of the diffeomorphism group. In consequence the formalism
of the theory is (by construction) insensitive to the very components distinguishing 
the cosmological spacetimes. On the other hand, recently, extending the standard LQC 
holonomy-flux algebra by all holonomies along piecewise analytic curves and 
imposing the symmetries on 
the classical level led to a viable embedding of LQC in LQG \cite{e-embed}.

To overcome the problems of $(i)$ another route [listed as point $(ii)$] was explored. There, 
instead of encoding the symmetries in the formalism one considered ``the operational 
approach'' -- defining the symmetries through relations between spacetime quantities 
provided by the LQG observables. The idea has been realized by constructing the 
coherent states peaked about symmetric spacetimes (see for example \cite{e-symm}). 
Provided the evolution of such coherent states in LQG can be controlled on the level 
of full theory this description can provide a definition of a (n effective) 
reduced formalism on its own. Such formalism will have much stronger foundation in it's 
LQG origin than LQC but applying it to probe for predictions is still a matter of 
the future. On the other hand, that formalism would loose its anchor in LQC, thus 
it would not be able to provide a solid connection with existing formulation of LQC 
or justification for the particular constructions implemented in it. As a consequence 
the utility of this approach as LQC test would be limited.

To mend this gap one needs to construct a precise dictionary between objects of 
LQG and LQC formalism, while keeping both the theories autonomous (the approach $(iii)$). 
Such dictionary should 
associate selected (possibly emergent) degrees of freedom of LQG with cosmological 
ones -- working as fundamental for LQC. In that aspect it is not necessary to restrict 
to symmetric or near symmetric states in LQG. Building this dictionary would be just 
a process of extracting some (global) degrees of freedom and as such should be well 
defined for \emph{any} quantum spacetime (or at least for a family of such which is 
sufficiently large to cover physically interesting scenarios). A good example of 
such procedure is a simple toy model used in \cite{awe-b1} to fix certain ambiguities 
in quantizing the Bianchi I spacetimes in LQC. There, as an example of the spin 
network one considered a regular (cubic) lattice of which edges have been excited to 
the first state above the ground one.
Much simpler models attempting to mimic cosmology, playing the role analogous to 
the one above, are also being constructed in Spin Foam formalism (often considered 
as the covariant formulation of LQG) \cite{rv-sfc1,*rv-sfc2,*hrv-sfc,*bdpg-un,*bfgl-un}.

It is also worth remembering, that outside of the \emph{top-down} approaches presented 
above there is a considerable literature on \emph{bottom-up} approaches, where the 
components of LQC are cast onto LQG structures. An example of such is the so called 
\emph{lattice LQC} \cite{we-lat}, where the structure of degrees of freedom and 
elementary operators inherent to perturbative cosmology are defined on the regular 
lattice, again playing the role of an example of an LQG spin network. Another, bit 
more distant example is testing the BKL conjecture in context of LQC \cite{ahs-bkl}.

The construction presented here is strongly motivated by the Bianchi I toy model 
mentioned two paragraphs above. it shares with this model the choice of spin network 
topology. Outside of this initial choice however 
we will keep the construction as general as possible (avoiding the dependence on 
particular prescription in either LQG or LQC) at the same time keeping full control 
over the assumptions entering the construction. In particular, as the unreduced side 
of the interface we will use \emph{genuine} LQG without any simplifications.
The dictionary will be provided by 
objects having a precise physical interpretation and well defined in both theories.

To have such a simple tool available becomes recently more than just a convenience, 
as there is an increasing amount of effort towards making the preliminary dynamical 
predictions of either conservative LQG \cite{t-qsd3,t-qsd5}, its modifications 
\cite{gt-aqg1,gt-aqg2,*gt-aqg3} or simplifications (in particular the simplifications 
of $SU(2)$ gauge to $U(1)^3$ \cite{acr-red,ac-red,*ac-let,b-b1}) via 
semiclassical approximations. One can thus confront the results of these projects 
with the predictions of LQC to either verify or falsify the latter.

Before proceeding with the construction of the LQG-LQC interface let us briefly
recall those element of both theories which will be relevant for its construction.

\section{Elements of LQC and LQC}

Since for all practical purposes both LQG and LQC are independent theories, just 
sharing common quantization methodology \cite{almmt-diffinv,abl-lqc} we proceed 
with presenting them separately. Let us start with LQG.

\subsection{Loop quantum gravity}

LQG is a quantization of canonical general relativity, owing a lot of its mathematical 
components to Young-Mills theories on the lattice. It's starting point is a $3+1$ 
canonical splitting, with the phase space coordinatized by the Ashtekar-Barbero 
variables: $su(2)$ valued connection $A^i_a$ and the densitized triad $E^a_i$, where 
$A^i_a$ is a combination of the Levi-Civita connection and exterior curvature 
$A^i_a = \Gamma^i_a + \gamma K^i_a$ (with $\gamma$ being the Barbero-Immirzi 
parameter) \cite{b-var}. 
As any representation of GR it is a constrained theory with the algebra of constraints 
generated by: the Gauss constraint, the spatial diffeomorphisms and the Hamiltonian 
constraint. To deal with them the Dirac program is implemented: theory is quantized 
without constraints (so called kinematical level), which are next solved on the 
quantum level (with solutions forming the \emph{physical} sector of the theory). 

The basic objects of the theory are the holonomies of $A$ along the 
piecewise analytic curves $U_\gamma(A) = \POrd \exp(\int_{\gamma}A_a^i\tau_i\rd x^a)$ 
and the fluxes of $E^a_i$ across surfaces $K^i = \int_S E^{ai}\rd\sigma_a$. 
Together they form form the holonomy-flux algebra, which is the fundamental 
object in constructing quantum theory. An application of the GNS (Gelfand-Naimark-Segal) 
construction to this algebra leads to the kinematical Hilbert space $\Hil_{\kin}^{\LQG}$ 
spanned by the cylindrical functions 
supported on the graphs embedded in $3$-dimensional differential manifold. These 
functions are conveniently labeled by the $su(2)$ representations (on each edge of 
the graph + the internal edges within the graph vertices) -- enumerated by half-integers. 
The quantum representation of holonomy-flux algebra provided by this construction is 
unique \cite{lost-uniq,*f-uniq}. This space is non-separable, and defining a separable 
physical Hilbert space structure in the further steps of Dirac program requires 
a nontrivial effort (see for example the discussion in \cite{bpv-osc2} and references 
therein).

On $\Hil_{\kin}^{\LQG}$ the constraints are solved in 
hierarchy (in the order: Gauss, diffeomorphism, Hamiltonian). The Gauss constraints 
distinguishes (through the kernel of an operator corresponding to it) a subspace by 
selecting out the \emph{gauge invariant} kinematical basis elements. These elements 
are characterized by the restrictions on 
the representation labels on the edges converging on a vertex (for each vertex of the 
graph), which restrictions can be thought of as analogs of the angular momentum addition 
rules in quantum mechanics.

Next the diffeomorphism constraint is solved by procedure of averaging \cite{almmt-diffinv} 
over group of finite diffeomorphisms, which act on the basis elements by modifying the 
embedding of the graphs supporting them, but without modifying the topology of the 
graphs or their quantum labels. In particular, in case the framework used involves 
a single particular graph the group averaging would simply lift the graph from embedded 
to the abstract one (see for example \cite{gt-aqg1}).

% observables 
The diffeomorphism-invariant Hilbert space $\Hil_{\diff}^{\LQG}$ provided by this procedure 
serves next as a main ``kinematical'' space, on which the (diffeomorphism invariant) 
observables are defined and Hamiltonian constraint is solved. Action of many geometry 
observables is explicitly known. An interesting property of many of them, among others 
the area, volume, angle or length operators is that their spectra are purely discrete, 
composed of (generically) isolated points. In this meaning it is often stated, that in LQG 
the space(time) is discrete. These operators are however not physical observables since
one more constraint remains -- the Hamiltonian one.

% the dynamics
In the last step of Dirac program one identifies the physical Hilbert space as a 
kernel of the Hamiltonian constraint operator and builds the physical observables 
out of the diffeomorphism-invariant kinematical ones. The latter is achieved through
the so called \emph{partial observable} framework \cite{d-obs,*r-obs}. In solving the 
Hamiltonian constraint several approaches are explored. Among them two approaches 
are considered as the most promising: \emph{the Master program} \cite{dt-master1,%
*dt-master2,*dt-master3,*dt-master4,*dt-master5} 
and the matter deparametrization. In the first approach, to avoid mathematical 
complications related with the structure of the constraint algebra one constructs a 
single non-negative definite operator out of all the constraints -- the so called 
master constraint. Then the physical Hilbert space is again given as a kernel of 
this operator. The last step was however not completed due to complicated mathematical 
structure of the constraint. 

In the second approach one uses the matter reference frames to provide (missing in 
the formalism) time variable \cite{gt-aqg4,*dgkl-scalar,hp-lqg-prl,*s-comm,*gt-matref}. 
This allows to reformulate the (originally constrained) theory as the free theory with 
a true Hamiltonian and where the original diffeomorphism-invariant Hilbert space 
itself or its proper subspace (depending on matter frame used) becomes the physical 
Hilbert space. The Hamiltonian is either former Hamiltonian constraint operator 
\cite{hp-lqg-prl,*s-comm,*gt-matref}
or its square root \cite{gt-aqg4,*dgkl-scalar}. In the former case its action on
$\Hil_{\phy}^{\LQG}\equiv\Hil_{\diff}^{\LQG}$ is explicitly known and feasible to 
compute \cite{hp-mg13}. In this formulation, 
the diffeomorphism-invariant kinematical observables become the physical ones. 

Among the components of the theory two particular objects will play the crucial 
role in constructing our dictionary. These are, the diffeomorphism-invariant area 
operator and the Euclidean part of the Hamiltonian constraint. let us focus our 
attention on the area first.

\subsubsection{\label{sec:lqg-ar}The area operator}

The properties of this operator are known in detail (see for example 
\cite{t-lqg,al-lqg-rev}). Its action on $\Hil_{\diff}^{\LQG}$ (i.e. on its spin network 
basis elements) is relatively simple \cite{flr-area}: the area of chosen (arbitrary) 
$2$-surface $S$ depends only on the $j$-labels of the edges of a spin-network 
reaching (or intersecting) this surface
\begin{equation}\label{eq:area-lqg}
	\Ar(S)\Psi[A] = 4\pi\gamma\lPl^2 \left[
		\sum_{e^+} \sqrt{j_{e^+}(j_{e^+}+1)} +
		\sum_{e^-} \sqrt{j_{e^-}(j_{e^-}+1)} \right] ,
\end{equation}
where $e^+$ are the incoming edges of the graph supporting $\Psi[A]$ which terminate 
on the surface, $e^-$ are the ones starting at the surface. $j_e^{\pm}$ are their 
respective $su(2)$ representation labels. 
The edges intersecting (piercing) the surface are counted as both incoming and 
outgoing (i.e. in that case a trivial $2$-valent node is temporarily introduced on the surface), 
thus their contribution is twice the terminating ones.  
The form of \eqref{eq:area-lqg} immediately implies, that the spectrum 
of $\Ar(S)$ is discrete. In particular the first non-zero value of the area is isolated 
from zero and equals
\begin{equation}\label{eq:lqg-area-min}
	A_1 = 2\sqrt{3}\pi \gamma \lPl^2.
\end{equation}

\subsubsection{\label{sec:H-euclid}The Euclidean part of Hamiltonian}

Classically the Hamiltonian constraint (or, more precisely, its part 
corresponding to gravity) is of the form
\begin{equation}
	C = \int \rd^3 x \sqrt{-q}\, \mathcal{C} , \qquad
	\mathcal{C} = \frac{\gamma^2}{2\sqrt{\det E}} E^a_i E^b_j 
		[ \epsilon^{ij}{}_k F^k_{ab} + 2(1-\gamma^2) K^i_{[a} K^j_{b]} ]
\end{equation}
where the field strength $F_{ab}$ is the curvature of the connection $A$.
The first term in $\mathcal{C}$ is the so called \emph{Euclidean term} of the 
Hamiltonian density. The loop quantization procedure, requires to express $\mathcal{C}$ 
in terms of the holonomies and fluxes -- the process known as Thiemann regularization 
\cite{t-qsd3}. In particular to express the curvature term $F$ one implements the known 
classical identity
\begin{equation}\label{eq:F-approx}
	F^i_{ab}X^a Y^b(x) 
	= \lim_{\Ar(\triangle(x))\to 0} \frac{U_{\triangle(x)}-1}{\Ar(\triangle(x))}
\end{equation}
where $\triangle(x)$ is the closed, piecewise analytic loop such that the vectors 
$X, Y$ are tangent to it at the point $x$, $U_{\triangle(x)}$ is the holonomy along 
this loop and $\Ar(\triangle(x))$ is the physical area of the loop. The right-hand 
side of this identity can be quantized directly in LQG, as the operators corresponding 
to the holonomy and the area are well defined.

The particular implementation of this identity (and the action of the resulting 
regularized $\hat{F}$ operator) depends on the specific construction (or prescription) 
of the Hamiltonian constraint. In the original construction by Thiemann \cite{t-qsd3} 
the (Euclidean part of) Hamiltonian constraint added at the vertices the spin network
the small triangular loops via introducing on the edges converging to it two $3$-valent 
nodes connected by an edge labeled by fundamental $su(2)$ representation ($j=1/2$).
Then the limit of shrinking this triangular loop to a point (at the node) was taken 
in the sense of the embedding, which nonetheless led to a diffeomorphism invariant 
result due to the nature of operator components in the approximation of $F$ (for 
details, see \cite{t-qsd5}).

In the alternative construction (see for example \cite{gt-aqg1}), where Hamiltonian 
constraint does not generate new edges, the loop has to be formed by existing edges 
of the spin network. It is defined by a requirement to form a plaquet -- the minimal 
closed surface, of interior not intersected by any edges. 
In particular, when the spin networks are supported on the regular lattice, these 
loops are the minimal squares. 

The particular construction of the Hamiltonian constraint also affects critically the 
structure of the physical Hilbert space. Below we will briefly discuss the issues 
related to it.

\subsubsection{\label{sec:Hphys}Physical Hilbert space structure}

Since in the Master program $\Hil_{\phy}^{\LQG}$ is defined only abstractly (formally)
to probe its properties we will focus on the deparametrization picture. 
To start with, we note that the kinematical Hilbert space which arises from the GNS 
construction is non-separable. The reason behind it is, that the induced inner product 
on $\Hil_{\kin}^{\LQG}$ makes the states supported on disjoint graphs orthogonal.
Each subspace of states supported on chosen graph is separable, with a discrete 
inner product, however the complete $\Hil_{\kin}^{\LQG}$ contains a continuum of 
(disjoint) graphs. Unfortunately, the gauge invariant and diffeomorphism invariant 
Hilbert spaces retain this property: The former puts only the restrictions 
on the labels of the graph without significantly decreasing the possible graph 
structures. The latter still allows for the continuum of distinct (orthogonal) 
graphs with a discrete inner product between them. Since in the deparametrization 
picture the space $\Hil_{\diff}^{\LQG}$ 
becomes the physical one, this deficiency is transmitted directly to the physical 
sector. As the non-separability can affect significantly the construction 
and properties of the coherent states and the statistical ensembles (see \cite{bpv-osc1}
for discussion of these issues on a simple quantum-mechanical example), this problem 
requires certain amount of care.

The particular treatment depends on whether the action of the (true in the 
deparametrization picture) Hamiltonian changes the graph topology. If the graph 
is fixed (see for example \cite{gt-aqg1}) the Hamiltonian distinguishes subspaces
invariant with respect to its action (supported on an unchanging graph). 
If the relevant observable operators are 
defined carefully and also preserve these subspaces, they become the superselection 
sectors, each of them being separable (as supported on one specific graph). The standard 
treatment calls then for a restriction of the studies to just one such sector.

If the Hamiltonian is graph changing, this procedure becomes less straightforward, 
although often superselection sectors can be distinguished due to the fact, that
(in specific prescriptions) the Hamiltonian changes the graph in a specific controlled 
way. This happens for example in case of the original construction of \cite{t-qsd3}. 
However, those superselection sectors can become already non-separable.

On the other hand, our experience from LQC shows (see the discussion in \cite{bpv-osc2}), 
that for certain models such restriction might be insufficient to provide a sufficiently 
large semiclassical sector reproducing General Relativity dynamics in small gravitational 
field regimes. In that case an alternative construction may be needed. 
Such alternative is provided for example in \cite{bpv-osc2}\footnote{The construction 
	there is presented on the example of the simple quantum mechanical system -- 
	an harmonic oscillator, however the applications to LQG are also discussed there.
}. There one makes use of available Lebesgue\footnote{The construction can also 
	be extended to many cases with a singular measure.
} measure on the space of superselection sectors. Then the inner product is defined 
as the integral with respect to that measure of inner products 
$\langle\cdot|\cdot\rangle_{\epsilon}$ on the single superselection spaces 
$\Hil_{\epsilon}$ (with $\epsilon$ being an abstract superselection sector label)
\begin{equation}\label{eq:Hinteg}
	\forall \psi,\phi\in\Hil:\ \langle\psi|\phi\rangle = \int \rd\mu(\epsilon)
		\langle\psi_{\epsilon}|\phi_{\epsilon}\rangle_{\epsilon} , 
\end{equation}
where $\psi_{\epsilon}, \phi_{\epsilon} \in \Hil_{\epsilon}$ are the restrictions 
(projections)
of the states to the single sector\footnote{This projections are known in the literature 
	as the so called \emph{shadow states}.
}. Action of the operators preserving the sectors 
extends in a straightforward way. The integral Hilbert space $\Hil$ is again separable.

\subsection{\label{sec:lqc}Loop quantum cosmology}

LQC, even when applied to the description of the inhomogeneous spacetimes, always relies 
on the reorganization of the geometry and matter degrees of freedom onto the quasi-global 
ones, for example the Fourier or spherical harmonic modes of the inhomogeneities/%
gravitational waves/matter (see \cite{mbgmm-G1,fmmmo-pert1}). As a consequence in 
this description there are always distinguished degrees of freedom corresponding to 
the ``background'' homogeneous spacetime. This distinction is achieved by partial 
gauge fixing, which is naturally distinguished in the cases of homogeneous spacetimes 
and in perturbative approaches. The remaining (inhomogeneous) degrees of freedom are 
then treated as the objects ``living'' on that homogeneous background on the equal 
footing with the matter fields. Thus, in all of 
the models a proper handling of the homogeneous spacetimes is an essential first step. 

Here, we focus on the simplest model representing such spacetime -- the model of 
Bianchi I universe. For the most of the paper we further fix the topology of its spatial slices 
to $3$-torus. The precise mathematical formulation of the LQC quantization of 
this model has been presented in \cite{abl-lqc} (isotropic spacetimes) and \cite{awe-b1} 
(actual quantization of the model following the procedures of \cite{abl-lqc}). It 
is performed via direct repetition of the procedure developed for LQG, although here 
the symmetries distinguish additional structure, which plays an essential role in 
the process. 

% natural gauge
First, the homogeneity distinguishes the natural partial gauge, in which the spacetime 
metric takes the form
\begin{equation}\label{eq:g-b1}
	g = -N^2(t)\rd t^2 + a_1^2(t)\rd x + a_2^2(t) \rd y + a_3^2(t) \rd z
\end{equation}
where $N(t)$ is the lapse function, $(a_1, a_2, a_3)$ are the scale factors in 
three orthogonal directions (in which the metric is diagonal) and $\oq = \rd x^2 %
+ \rd y^2 + \rd z^2$ is the isotropic fiducial metric constant in comoving coordinates 
$(x, y, z)$. This choice fixes all the spacetime diffeomorphisms up to: a global time 
reparametrization, and the (rigid in $\oq$ metric) global spatial translations. 

% variables and h-f algebra
Similarly to the general GR case, we select the Ashtekar-Barbero variables, although 
here the fiducial metric distinguishes the orthonormal triad $\ofe^a_i$ of vectors 
pointing in eigendirections of the physical metric and preserving the spatial symmetries
of the system. That structure again allows to partially gauge-fix the variables 
through selecting
\begin{equation}\label{eq:AE-lqc}
	A_i^a = c^i (L_i)^{-1} {}^o\!\omega^i_a ,
	\qquad
	E^i_a = p_i L_i V_o^{-1} \sqrt{\oq} \ofe^a_i ,
\end{equation}
where ${}^o\!\omega^i_a$ is a co-triad dual to $\ofe^a_i$, $V_o$ is the fiducial
(with respect to $\oq$) volume of the homogeneous spatial slices and $L_i$ are their 
(also fiducial) linear dimensions. 
The global coefficients $c^i$ and $p_i$ are the so called connection and triad 
coefficients. They form the canonical set with Poisson bracket $\{c^i,p_j\} = %
8\pi G\gamma\delta^i_j$.  

%h-f algebra
In the next step one constructs the holonomy-flux algebra. Here however one notices, 
that upon the choice \eqref{eq:AE-lqc} one can just select the subalgebra of 
holonomies $U_i^{(\lambda)}$ along the straight edges in direction $\ofe_i$ and the 
fluxes $S_i$ along the unit squares orthogonal to $\ofe_i$ as they suffice to separate 
the phase space points. Further, the fluxes $S_i$ can be associated with the triad 
coefficients themselves as 
\begin{equation}\label{eq:S-p}
	S_i = p_i. 
\end{equation}

On such restricted (subalgebra of the) holonomy-flux algebra one implements the 
GNS construction, arriving to the unique quantum representation \cite{a-notes,*ac-uniq}. 
The kinematical Hilbert space resulting from this construction is a product of 
the square summable functions on the Bohr compactification of the real line
\begin{equation}
	\Hil_{\kin}^{\LQC} = \left[ \Sigma^2(\bar{\re}_{\Bohr}, \rd\mu_{\Bohr}) \right]^3 ,
\end{equation}
each copy of the space corresponding to one direction of $\ofe_i^a$.

The basic operators --quantum counterparts of the holonomy-flux algebra elements-- 
are the holonomy operators $\hat{U}_i^{(\lambda)}$ and the unit flux 
operators $\hat{p}^i = \hat{S}^i$. the latter are known in the literature as 
the \emph{LQC triad operators} due to the simple classical relation \eqref{eq:S-p}. 
We will implement the same naming policy here. One has to remember however, that 
these operators represent \emph{the fluxes}, not the triads. As in full LQG, 
in LQC the operators 
corresponding to the holonomies and triads themselves \emph{do not exist}. In fact,
the relation between $p^i$ the scale factors $a_i$ in \eqref{eq:g-b1} (see \cite{awe-b1})
shows, that the operators $\hat{p}^i$ measure \emph{the area of the maximal surface 
orthogonal to $\ofe_i^a$}.

The kinematical states $\ket{\psi} \in \Hil_{\kin}^{\LQC}$ automatically satisfy 
the Gauss and diffeomorphism constraints. The only gauge transformations left after 
the partial gauge fixing -- the global spatial translations act on the elements 
of $\Hil_{\kin}^{\LQG}$ as an identity. The only nontrivial constraint remaining 
is the Hamiltonian one.

To construct the operator representing the (gravitational term of the) Hamiltonian 
constraint one repeats the Thiemann construction of LQG, partially discussed in 
sec.~\ref{sec:H-euclid}. For the Bianchi I model, the Lorentzian (exterior curvature)
term is proportional to the Euclidean one, thus only the quantization of the latter 
is needed. For that one again has to deal with the field strength term, which is 
approximated by holonomies along a closed loop via \eqref{eq:F-approx}. Since only
the holonomies along the diagonal directions of $q$ are available the loop is a 
rectangle oriented in directions of $\ofe^a_i$.

Here however 
we see an important difference with respect to LQG. In the full theory either 
(depending on the formulation) one could move (shrink) the loop in the embedding 
manifold and the transformation has not modified the physical area of the loop.
In LQC, the presence of the \emph{background fiducial metric} $\oq$ --an object 
responsible for the rigid relation between $a_i$ and $p_i$-- fixes a unique 
relation between the embedding (fiducial) area of the loop and its \emph{physical area}.
For a loop spanned by (holonomies along) vectors $\ofe_j$, $\ofe_k$ we have
\begin{equation}
	\Ar(\square_{jk}) = \epsilon^{i}{}_{jk} p_i \lambda_j\lambda_k
\end{equation}
where $\lambda_i$ are the fiducial lengths of the straight edges in direction $\ofe^a_i$ 
forming the loop. 

This relation plays a crucial role in fixing the action of the Hamiltonian constraint 
in LQC \cite{aps-imp}. In principle in LQC one can set the fiducial edge lengths 
$\lambda_i$ freely which would allow to construct the loop or arbitrarily small areas. 
On the other hand in LQG the spectrum of the area operator is purely discrete and 
the first non-zero value of the area is determined by the theory. Since the goal of 
formulating LQC was the construction of the simplified settings approximating or mimicking 
the full theory as close as possible, this particular property (discreteness of the area)
has been \emph{parachuted from LQG}. The fiducial lengths $\lambda_i$ are fixed by the 
requirement that the physical area of the loop equals
\begin{equation}\label{eq:lqc-area}
	\Ar(\square) =: \Delta = 2 A_1 ,
\end{equation}
where $A_1$ is provided by \eqref{eq:lqg-area-min}.
Loop of these dimensions is then considered as the minimal loop realized in LQC. 
The reason, why one takes as the minimal area $2A_1$ instead of $A_1$ \cite{awe-cov} 
is that it should correspond to the area of the surface \emph{pierced} by the edge, 
not just with the edge terminating on it. This requirement follows from the 
semi-heuristic lattice construction in \cite{awe-b1} and will became apparent 
in the process of constructing the dictionary in further sections.

The particular way in which the requirement \eqref{eq:lqc-area} fixes the lengths 
$\lambda_i$ is construction dependent and in the past led to several distinct 
prescriptions in quantizing the Bianchi I model in LQC (see for example \cite{ch-b1}). 
Subsequently one choice has been distinguished by the construction in\cite{awe-b1} 
and by certain invariance requirements in application of the framework to the 
noncompact universes. At present the construction introduced in \cite{awe-b1} 
is considered to be the unique consistent prescription.

With the lengths of holonomies fixed and the remaining components in the Hamiltonian 
constraint regularized via Thiemann construction, one arrives to the Hamiltonian 
constraint operator, which is a difference operator acting on the domain of 
elements of $\Hil_{\kin}^{\LQC}$ supported on the finite number of points 
$\ket{p_1,p_2,p_3}$ 
[see eq.~$(3.35)$-$(3.37)$ in \cite{awe-b1}]\footnote{For technical reasons (simplicity 
of the constraints) in \cite{awe-b1} different labeling of the kinematical basis 
	states is used.
}. Given that, after coupling with appropriate matter fields the dynamical sector 
of the theory is determined by either group averaging \cite{almmt-gave,*m-gave1,%
*m-gave2,*m-gave3,*m-gave4} (see also \cite{aps-det,klp-gave} for applications 
in context of LQC) and partial observable formalism or via deparametrization. 

Here however we encounter the same problem as in full LQG: the kinematical Hilbert 
space $\Hil_{\kin}^{\LQC}$ is non-separable (due to the discrete inner product).
Since in the deparametrization picture it becomes the physical space, the latter 
is also non separable. In isotropic LQC the procedure of dealing with this problem
makes use of the fact, that the Hamiltonian (or Hamiltonian constraint) distinguishes
certain subsets (``lattices'') invariant under its action. The subspaces of states 
supported on those sets form then superselection sectors, each being separable. 
Subsequently one choses just one sector to describe the dynamics.

An extension of this approach to the anisotropic LQC is nontrivial, since for example 
in Bianchi I case the superselection sector ``lattices'' are formed out of families 
of sets dense on surfaces of codimension $1$ in the configuration space \cite{mbmmwe-b1}.
Furthermore, it is not at all obvious, that a single sector would admit a proper 
semiclassical regime, where low energy dynamics conforms to GR. 

In order to deal with such difficulties one can again implement the integral Hilbert 
space construction \cite{bpv-osc2} following \eqref{eq:Hinteg}. In the case of known 
isotropic models \cite{aps-imp,apsv-spher} the results following from implementing 
this construction are (up to minor corrections) equivalent to the ones provided 
by treatment involving just one superselection sector.

\section{\label{sec:dict}The dictionary}

The review in the previous section shows clearly, that both LQG and LQC describe 
the geometry via 
very distinct sets of degrees of freedom. The principal difference between these 
frameworks is the presence of background (fiducial) geometry in LQC and absence of 
such in LQG. This LQC background structure is distinguished by the symmetries of 
the theory on the classical level. By its very construction, LQG does not ``detect'' 
these symmetries as there the symmetry transformations are just a specific class 
of finite diffeomorphisms, to which the framework is insensitive. As a consequence, 
building an LQG state representing the cosmological spacetime is a nontrivial task 
\cite{e-symm}.

Our goal here is constructing the relation between the frameworks on the kinematical 
level, that is we do not address the matter of agreement of the dynamics between these 
two frameworks. Answering the question whether the LQC dynamics of a state representing 
a universe is a good approximation to the dynamics of the state representing the 
same universe in LQG, is beyond the scope of this article. Instead, we explore the 
interplay between the structures in both theories and the consequence the consistency 
requirements of one framework impose on the other. In the process we try to keep 
the full control over the 
initial assumptions entering the construction and to determine the freedom left after 
making these assumptions.

Our (principal) point of departure is the observation, that the basic quantities 
characterizing the state in LQC --the areas $p_i$ of maximal surfaces orthogonal 
to the basis triad vectors-- correspond to observable quantities well defined for 
any physical LQG state. To see that, let us fix the type of spacetime represented 
and the approach to constructing the dynamical sector.

\begin{itemize}
	\item Our objects of studies will be physical states corresponding to the (homogeneous 
		but not isotropic) Bianchi I universe, of which the spatial slices have a $3$-torus 
		topology (although the results can be easily generalized to the noncompact flat case).
		Thus the embedding manifold for the kinematical spin-networks will be topologically 
		$T^3$.
	\item Since the dictionary we are constructing involves the correspondence between the 
		geometry degrees of freedom only and does not employ the dynamics of the system, 
		we can safely assume, that it will not depend 
		on the type of matter content coupled to gravity. Thus we can work with chosen 
		particular type of matter and the results will automatically generalize to other 
		types of matter. For that purpose we select the construction of the 
		dynamical sector through the deparametrization with respect to the irrotational 
		pressureless dust serving as the time frame in both LQG \cite{hp-lqg-prl} and LQC 
		\cite{hp-lqc-let}.
\end{itemize}
 
While there is no consensus in the area as to whether the particular matter content 
selected above is 
well motivated physically, this choice provides simple and precise framework, 
circumventing the difficulties present in other approaches. In particular the 
diffeomorphism invariant Hilbert space in LQG and the kinematical Hilbert space 
of LQC automatically (and without any modifications or corrections) became the 
physical Hilbert spaces of their respective theories. Same applies to the area 
operators in LQG specified in sec.~\ref{sec:lqg-ar} and the ``triad'' operators
$\hat{p}^i$ in LQC -- they become physical observables. As a consequence they 
can be used directly, when comparing \emph{physical} areas.

% now relation between areas
Given that one can associate with each physical state $\ket{\Psi}\in\Hil_{\phy}^{\LQG}$
(where we do not require this state to be supported on just one spin network) a 
cosmological state $\ket{\Phi}\in\Hil_{\phy}^{\LQC}$ such that the expectation values
of the area operators along the ``flat and orthogonal'' maximal surfaces agree with 
the expectation values $\bra{\Phi}\hat{p}^i\ket{\Phi}$. Despite lack of background 
metric, the notion of flatness and orthogonality\footnote{We use these terms since for 
	homogeneous spacetimes (of diagonal spatial metric) they coincide with the standard 
	meaning of flatness and orthogonality. This agreement may however not extend beyond 
	that class of spacetimes.
} can be made precise in terms of 
the expectation values. Below we present one of possible constructions. it will not 
be used in construction of LQG-LQC interface and it is presented solely as an example 
that making the abovementioned association between LQG and LQC states is possible.
One of many ways to define such construction is:
\begin{enumerate}
	\item First one chooses on the embedding manifold a point $p$ and a triad of 
		vectors anchored on it.
	\item The angle operator in LQG is well defined \cite{t-lqg}, thus one can 
		distinguish (also in the embedding manifold) a triple of $2$-surfaces of $T^2$ 
		topology --sections of the embedding manifold-- such that the respective pairs 
		of distinguished triad vectors are tangent to them (at their intersection). 
		Their relative orientation 
		is then fixed by the requirement that the expectation value of the angle operator
		corresponds to normal angles.
	\item Finally, the ``flatness'' of the surfaces is enforced by requirement that 
		the surfaces minimize their physical area (again in terms of the expectation 
		values of the area operator).
\end{enumerate}
Then the area of the distinguished surfaces is associated with respective areas 
$\langle \hat{p}_i \rangle$ in LQC.

This association between LQG and LQC states is far from unique: $(i)$ it is fixed 
just by relation between expectation values of three observables, which is obviously 
insufficient to determine the state, $(ii)$ the areas of the distinguished surfaces 
may depend on the point $p$ and the chosen vector triad, and $(iii)$ there may not 
be a global minimum of the areas in the point $3.$ of the above construction.
At this moment however we do not look for the uniqueness of the association. We just 
want to show that it can be made. It may not be particularly useful for constructing 
the LQC limit of LQG, especially because so far we have not introduced any notion of 
symmetry. Literally, any (even very inhomogeneous) physical LQG state can be used 
in this construction.

On top of that deficiency, at present we are lacking any relation with the 
auxiliary structures in LQC, which is necessary to really understand the 
relation between the frameworks. Therefore in further studies we are going to restrict 
the space of possible physical states, by selecting off a very specific (yet sufficiently
large to accommodate the physically interesting spacetimes)
set of spin networks supporting the states.

\subsection{\label{sec:lat}The lattice spin-network}

Our construction is heavily inspired by the semi-heuristic construction introduced 
in \cite{awe-b1}. There, one equipped the embedding $\re^3$ manifold with a fiducial 
metric $\oq$ and used it to define a regular lattice in it. This lattice has been next used 
to construct a specific spin-network by associating with each edge (link) of the lattice 
a $j$-label ($j=1/2)$ corresponding to the fundamental $su(2)$ representation -- 
a minimal non-zero value allowed by the theory. Then, the gauge-invariant state has been 
distinguished as supported on this spin network only. 

Given that state, one could introduce a so called fiducial cell -- a compact region of 
space acting as the infrared regulator of the theory. It was chosen such that its edges 
were parallel (in the sense of $\oq$) to the edges of the lattice. Due to fixing of the 
$j$-labels the area of each face of the cell was then proportional to the number of the 
lattice edges piercing it. With these areas one subsequently associated the values of 
$p_i$ in LQC.

Given that association, the regularity of the lattice allowed in turn to associate with 
each plaquet (minimal square loop of the lattice) a ``physical''\footnote{On the formal 
	level this value cannot be associated with the expectation value of the LQG area operator 
	since no edge intersect such plaquet. It can be however made precise with use of the 
	so called \emph{dual graph} -- a technique often applied in the spin-foam approaches.
} area. Finally the requirement that all these areas equal $\Delta$ \eqref{eq:lqc-area} 
fixed the fiducial lengths of the edges of the plaquets, which in this construction are
the curves along which the fundamental holonomies are taken. 

In this article we expand on this idea, dropping however most of the assumptions made  
in \cite{awe-b1}. 
%
% And now the construction
%
To start with, we consider a single spin network, embedded in the $T^3$ manifold 
defined above. We further assume that our spin-network is topologically equivalent 
to the regular lattice\footnote{To be mathematically precise, we define the graph 
	which admits a set of discrete symmetries of the regular (closed) lattice on $T^3$.
} or is a proper sub-graph of such. This graph is next equipped with the 
$su(2)$ $j$-labels on the graph edges (and internal edges at each node)\emph{where 
in particular the value $j=0$ is allowed.} If the original graph is the 
proper subgraph of the (topologically) regular lattice, it is completed to the lattice 
by adding appropriate edges with $j=0$ and appropriate vertices. We further  
assume that the lattice is minimal: since two edges of $j=0$ entering $2$-valent 
vertex can be always replaced with one single edge, given a lattice spin-network
we perform such reduction, whenever it does not destroy the regular lattice topology.
Such construction of a spin-network, although abstract instead of embedded is used 
for example to formulate the \emph{Algebraic Quantum Gravity} \cite{gt-aqg1} framework.

% LQC bacground
To introduce the cosmological background structure we note, that a large class of 
spatial diffeomorphic gauge fixings can be implemented via equipping the embedding 
manifold with a metric tensor. For every spin network one can define a fiducial
isotropic metric $\oq=\rd x^2 + \rd y^2 + \rd z^2$ such that ${\rm x}_i:=(x,y,z)$ 
are the functions defined along the edges of the graph, preserved by the discrete 
symmetries of the graph: for a cyclic permutation of the nodes along one lattice 
direction $i$\footnote{The \emph{direction} is defined here by topology of the graph: 
  each direction is the set of the classes of equivalence of minimal closed loops 
  not shrinkable to a point on the embedding manifold.
} the function $\rx_i(\vec{x})$ changes as follows 
\begin{equation}
	\rx_i \mapsto \rx_i+\lambda_i , \qquad \lambda_i := 1/n_i
\end{equation}
where $n_i$ is the number of graph edges forming a closed loop in direction $i$. 
The coordinates on the graph are next extended smoothly (non-uniquely) to the 
whole embedding manifold.

% invariance
It is worth reiterating, that neither the partial gauge fixing introduced here
nor the auxiliary structure play any role in describing the physics. All the physical 
geometry observables are insensitive to this choice, as their action depends only 
on the topology of the graph and its quantum labels.

% facets and the remaining gauge freedom
Given the regular lattice defined above, one can precisely implement the construction 
of the LQG$\leftrightarrow$LQC dictionary
specified at the beginning of sec.~\ref{sec:dict}. For that we choose the constancy 
surfaces $S_i$ of the coordinates $\rx_i$ (thus orthogonal to $\ofe_i^a$). The areas 
of these surfaces (expectation value of the operator \eqref{eq:area-lqg}) are then 
associated with LQC ``triad'' or ``flux'' coeffcients 
\begin{equation}
	\langle\Ar(S_i)\rangle = p_i := \langle\hat{p}_i\rangle .
\end{equation}
The form of \eqref{eq:area-lqg} implies immediately, that these areas do not depend 
on the way the coordinates $\rx_i$ have been completed between the elements of 
the graph. The values of $p_i$ depend however on the $j$-labels of the edges intersecting, 
terminating and contained within $S_i$ which may differ depending on the particular 
choice of the surface.\footnote{We remind that no restriction is made on the distribution 
  of the values of $j$-labels on the graph.
} Thus, this association is not unique. To emphasize this fact, we further denote these 
values as $p_{i,\rx_i}$ and the surfaces themselves as $S_{i,\rx_i}$.
At this point we have to remember however, that there is some residual diffeomorphism 
freedom left in the system: the rigid (with respect to $\oq$) translations in $\rx_i$.
We will exploit this freedom in the next subsection to complete the dictionary 
construction.

% Here comment on the dynamics
Before doing so however, we have to address one issue: since the dictionary will 
rely on the auxiliary structure, it is critical to check how it will be affected 
by the dynamics.

In case, the LQG Hamiltonian is graph-preserving (like for example in \cite{gt-aqg1})
there is no problem: the only elements affected are the spin labels. The graph itself 
does not change, thus its embedding in the manifold can be assumed to be constant in 
time. This in turn allows to keep the auxiliary structure constant.

The situation complicates a bit when the Hamiltonian is \emph{graph changing}. There 
the preservation of the structure of the graph depends on particular form of that 
Hamiltonian. For example in original canonical LQG construction of \cite{t-qsd3} the 
Hamiltonian adds edges with $j=1/2$ labels forming a triangular loops with existing 
ones. This can be easily implemented in the construction considered here, if instead 
of triangle we add a square loop with two new $j=1/2$ edges. The new spin network 
can be then easily completed to a (topologically) regular lattice by adding $j=0$ edges. 
The auxiliary structure can then be easily rebuilt, essentially in tow ways:
\begin{enumerate}
	\item The coordinates $x_i$ and fiducial metric can be redefined so the new lattice 
		becomes regular in them. This corresponds to discontinuously shifting the vertices 
		of the graph to new positions on the embedding manifold (passive diffeomorphism).
		The discontinuity is however not a problem, as the Hamiltonian flow is not used 
		in the construction of the dictionary and the auxiliary structure can be defined 
		at each time slice independently.
	\item \label{it:latex-dense}
		The new nodes can be placed in the center (with respect to $\oq$ of existing 
		plaquets. Then, the lattice would be a particular realization of the \emph{dense 
		spin-network} \cite{ag-dense}. It would however loose the regularity. The latter 
		could in principle pose a problem as the fiducial length of the edges will be 
		an essential component of the dictionary. In that case however one should average 
		any quantity evaluated on the graph over the diffeomorphisms changing the 
		fiducial lengths of the edges but preserving the directions (with respect to 
		metric $\oq$) of the edges of the graph. For quantities which are averages 
		weighted by the fiducial lengths this averaging procedure yields exactly the 
		same results as the ``uniformization'' defined in the point above 
		(see Appendix~\ref{app:stat}).
\end{enumerate}

It is important to note that the regularity assumption can be replaced with the 
average over (passive) diffeomorphisms preserving the directions of the edges 
(with respect to $\oq$). Indeed one can represent the (relevant for the graph) 
passive diffeomorphisms as random distributions of the vertices coordinates over 
the interval $[0,1]$, see Appendix~\ref{app:stat}.

\subsection{\label{sec:avg}The averaging procedure}

At present a relevant difference remains between the LQG state constructed previously 
and its LQC analog. The LQC state is constructed with implicit assumption of representing 
the homogeneous spacetime, whereas the LQG one can a priori be highly inhomogeneous. 
This implies that the association of the values of $p_i$ has to involve some kind of 
averaging (over the inhomogeneities) procedure. In LQG the lack of background structure 
makes the definition of such averaging difficult. Here however the choice of the spin 
network graph and partial gauge fixing allowed to construct the necessary background 
structure.

To employ it, we now consider the remaining rigid translations as the \emph{active 
transformations} shifting the surfaces $S_{i,\rx_i}$ along the graph and define the 
values $p_i$ as the averages with respect to this translation group. In the mathematically 
precise sense the variables $p_i$ are chosen to equal the expectation values of the 
area operator (of each surface $S_i$) \emph{averaged over the rigid spatial translation 
group}. A simple calculation using \eqref{eq:area-lqg} shows then, that
\begin{equation}\label{eq:p-dict}
	p_i = \frac{8\pi\gamma\lPl^2}{n_i} \sum_{e\in \{e\}_i} \sqrt{j_e(j_e+1)} 
	=: \frac{8\pi\gamma\lPl^2}{n_i} \Sigma_i,
\end{equation}
where $\{e\}_i$ is the set of all the edges of the graph which point in direction of 
$\ofe_i$ and $n_i:=1/\lambda_i$ is the number of graph edges pointing in direction 
of $\ofe_i$ and forming a close loop. Here, the edges terminating on the surface or 
contained within it do not contribute, as that would require tuning the translations
and such translations are a zero measure set within the distinguished translation 
group.

% plaquet
Next component needed in our dictionary are the (physical) areas of the minimal 
loops (plaquets) needed to approximate the field strength operator. To evaluate 
these areas we proceed exactly as in the case of the surfaces $S_i$: we average 
the relevant area operators over the rigid (active) translations group. Again, 
a simple calculation yields (here we denote these areas by $\sigma^i$)
\begin{equation}\label{eq:sigma1}
	\sigma_i = \frac{8\pi\gamma\lPl^2}{n_1 n_2 n_3} \sum_{e\in \{e\}_i} \sqrt{j_e(j_e+1)}
	= p_i \frac{\lambda_i}{\lambda_1\lambda_2\lambda_3} .
\end{equation}
As a consequence, the ratio $\sigma_i/p_i$ (no summation over $i$) 
does not depend on the $j$-labels of the spin network. It depends 
\emph{only on the number of edges of the graph} which is an expected result since 
(once the averaging is implemented) each surface $S_i$ can be simply composed 
out of $n_1n_2n_3/n_i$ plaquets. It is important to note, that, even though we 
are permitting the edges with $j=0$, the numbers $n_j$ are invariant due to gauge 
invariance and the requirement that the lattice is minimal.

% standard LQC treatment and consequences
The standard procedure implemented in LQC would call in the next step for an 
association $\sigma_i=\Delta$, where $\Delta$ is defined via \eqref{eq:lqc-area}.
This would fix $\lambda_i$ as the functions of the phase space, leading exactly 
to the dependence found in \cite{awe-b1}. Here however we do not implement this step,
expecting in turn that the value of $\sigma_i$ should follow from the properties 
of the spin network. Thus, the only property at our disposal is the relation of 
$\sigma_i$ with the average (over the graph) of $j$-labels associated with edges 
in direction $\ofe_i$
\begin{equation}\label{eq:sigma-inv}
	\sigma_i = 8\pi\gamma\lPl^2 \overline{[ \sqrt{j(j+1)} ]_{e_i}} =: \Delta_i,
\end{equation}
where the symbol $\overline{[ \cdot ]}$ denotes the average over the graph. For 
the models aimed to reproduce the cosmological spacetime via specific semiclassical 
states the value of such average depends on the details of the model and may 
in principle differ significantly from the value $\Delta_i = \sqrt{3/2}$ consistent 
with $\sigma_i=\Delta$ (see for example \cite{ac-red}). In such models, the values 
$\sigma_i$ do not need to be fixed by any fundamental constant and a priori may depend 
on the state.

\subsubsection{\label{sec:trans-alt}Alternative averaging procedure}

The results of the averaging procedure implemented above can be easily understood 
on the intuitive level if we introduce a convenient decomposition of the group 
of rigid translations. First, one can introduce a discrete group $\ig^3_{n_1,n_2,n_3}% 
:=\ig_{n_1}\times\ig_{n_2}\times\ig_{n_3}$ of cyclic permutations of the graph 
vertices. The quotient of the group of translations over $\ig^3_{n_1,n_2,n_3}$ is the 
group of translations over the distances $\delta_i\in [0,\lambda_i)$. The averaging
procedure can now be split onto two steps:
\begin{itemize}
	\item Averaging over $\ig^3_{n_1,n_2,n_3}$, which simply replaces the $j$-label
		of a single edge with the average $\overline{[j]}_{e_i}$ of all the edges parallel 
		to it.
	\item Averaging over the quotient group. The result of this step follows directly 
		from the observation that upon the action of the quotient group (with exception 
		of the group neutral element which forms a zero measure set) each plaquet of 
		the graph is intersected by exactly one edge (now carrying the averaged $j$) 
		orthogonal to it.
\end{itemize}
The above procedure leads immediately to \eqref{eq:sigma-inv} and \eqref{eq:sigma1} 
and further, after reassembling the surfaces $S_i$ out of the plaquets, to 
\eqref{eq:p-dict}.

\bigskip

% why we need an extension (we don't)
The potential application of the studies performed here to the models, where the 
cosmological spacetime is defined by the semiclassical (often chosen to be coherent) 
state, lead to another complication. So far we have considered the single graph.
Why such choice is perfectly fine to define a basis of a Hilbert (sub)space, it may 
be insufficient for such models. Therefore one needs to extend the dictionary to 
incorporate large number of such spin-network ``superselection sectors''. We provide 
such extension below, using the integral Hilbert space construction presented in 
sec~\ref{sec:Hphys}.

\subsection{\label{sec:int}The integral extension}

As in the case of a single lattice space, here we are going to define some subspace 
of $\Hil_{\phy}^{\LQG}$. 
\begin{enumerate}
	\item We start with a single lattice spin network defined in sec.~\ref{sec:lat} 
		(without introducing the background metric $\oq$).
	\item \label{it:max_s} The plaquets of the spin network define three classes 
		of surfaces of topology
		$T^2$ (maximal surfaces on the embedding manifold).\footnote{One can introduce 
			a notion of parallel edges terminating in a $6$-valent node as the pair not 
			being the edge of a single plaquet and next build the surface by selecting 
			a plaquet and extending the surface by including plaquets whose least two 
			edges are parallel to edges contained already by the surface.
		} such that within one class the surfaces do not intersect each other and 
		all the intersections of the surfaces of distinct classes are of $S^1$ topology.
	\item \label{it:surf} The discrete classes of surfaces are next completed to congruences of 
		the embedding manifold (using the surfaces of the same topology) keeping the 
		requirement, that intersections between representants of different classes 
		are $S^1$. This (non-unique) extension always exists.
	\item We extend the original lattice spin network to a class of disjoint
		spin networks of which edges are intervals of the intersections of surfaces 
		defined above. This class is selected in such a way, that
		\begin{enumerate} 
			\item each point of the embedding manifold is a node of exactly one spin 
				network in the class, and
			\item given two graphs of the set, the maximal $T^2$ surfaces of point \ref{it:max_s}
			are interlaced, that is within each class of surfaces, between two surfaces 
			of one graph  there is exactly one surface of the other graph.
		\end{enumerate}
	\item Each spin network is completed to a (topologically) regular lattice 
		by adding edges with $j=0$.
\end{enumerate}
In its essence this method produces a continuum of (topologically) regular lattices
of which edges are ``parallel''. They define a distinguished coordinate system $\rx_i$
where the coordinates are functions constant on the $T^2$ surfaces from point \ref{it:surf}.
Given this coordinate system, one can equip the manifold with a background fiducial metric
$\oq = \sum_{i=1}^3 (\rd\rx_i)^2$. This construction is of course quite restrictive,
however it allows to preserve the well defined notion of principal (diagonal) directions
of the single lattice. 

One way to produce the specific example of such set of spin-networks is to start 
with one regular lattice, equip the embedding manifold with the fiducial metric $\oq$
(as in sec.~\ref{sec:lat} and then act with the active rigid translations defined in 
sec.~\ref{sec:avg}. The family of possible continuous sets defined in points $1.-5.$ 
is however much bigger. In particular, the lattices do not need to be regular with 
respect to the metric $\oq$.

An important property of the selected set of spin networks is that it admits a well 
defined Lebesgue measure $\rd\sigma$ induced by the Lebesgue measure of a minimal 
cube of any (chosen arbitrarily) lattice within the set. This measure can now be 
used to construct the integral Hilbert space via \eqref{eq:Hinteg} via setting 
$\rd\mu(\epsilon)=\rd\sigma$. Choosing different minimal cubes will lead to unitarily 
equivalent spaces.

% now averaged observables and averaged \lambda's
Given the new (integral) Hilbert (sub)space we proceed with defining averaged 
quantities $p^i$, $\sigma^i$ exactly as in sec.~\ref{sec:avg}. The only difference is
an additional integration over the selected set of lattices. The calculations yield
\begin{subequations}\begin{align}
	p_i &= \frac{8\pi\gamma\lPl^2}{n_i} \int \rd\sigma(\vec{\epsilon})
		\sum_{e\in \{e\}^i_{\Gamma(\vec{\epsilon})}} \sqrt{j_e(j_e+1)} 
	=: \frac{8\pi\gamma\lPl^2}{n_i} \overline{\Sigma}_i, 
	\\
	\sigma_i &= \frac{8\pi\gamma\lPl^2}{n_1 n_2 n_3}
	\int \rd\sigma(\vec{\epsilon}) \sum_{e\in \{e\}^i_{\Gamma(\vec{\epsilon})}}
	\sqrt{j_e(j_e+1)}
	= p_i \frac{\bar{\lambda}_i}{\bar{\lambda}_1\bar{\lambda}_2\bar{\lambda}_3} 
	= 8\pi\gamma\lPl^2 \overline{[ \sqrt{j(j+1)} ]}_{e_i,\vec{\epsilon}} =: \bar{\Delta}_i,
	\label{eq:avg-sigma}
\end{align}\end{subequations}
where $\vec{\epsilon}$ labels the superselection sectors and $\Gamma(\vec{\epsilon})$
is the graph corresponding to superselection sector $\vec{\epsilon}$. We see, that 
the quantities $\lambda_i$, $\Delta_j$ are now simply replaced by their averages 
$\bar{\lambda}_i$, $\bar{\Delta_j}$ over the superselection sectors. The quantities
$\bar{\lambda_i}$ are now \emph{defined} by \eqref{eq:avg-sigma} and do not necessary 
correspond to the average over fiducial lengths of the edges (which in turn may not 
be the constants of the graph). The average $j$-label, is however a proper average
\begin{equation}
	\bar{\Delta}_j = \int\rd\sigma(\vec{\epsilon})\Delta_j(\vec{\epsilon}) .
\end{equation}

The above result has been found under assumption of a specific compact topology 
of the embedding manifold. However in LQC a consistency restrictions that actually 
do fix the theory originate in models of the noncompact universes, some of them 
already in isotropic sector. 
Therefore it is prudent to extend our dictionary to such case and to incorporate in it 
the notion of isotropy. For simplicity we will consider single lattice states only, 
although the generalization to the integral states of sec.~\ref{sec:int} is not 
difficult.

\subsection{The noncompact extension and the isotropy} 

In order to keep the model as simple as possible we consider its extension to the 
flat space, assuming $R^3$ topology.

\subsubsection{The extension to $\re^3$}

In LQC the standard method of dealing with infinities due to Cauchy slice noncompactness
is selection of a compact region --the so called fiducial cell-- which then becomes 
the infrared regulator of the model. The physical predictions are then extracted within 
the regulator removal limit. Well definiteness of that limit is the first consistency 
condition imposed on LQC and is precisely the origin of the so called \emph{improved 
dynamics} prescription \cite{aps-imp}. Here we follow the same idea: We start with the 
construction of a single lattice spin network, as specified in sec.~\ref{sec:lat}, 
although now the lattice is open and infinite. We then introduce the background structure
exactly as in the compact case and distinguish the regulator -- a rectangular cube 
of edges pointing in directions of $\ofe_i$. The expansion of the regulator is well defined 
in terms of the number of edges forming the interval of ``straight lines'' which is 
contained within the cube (or ,equivalently, the number of elementary cells stacked 
along the edge of the cube).

The remaining spatial gauge freedom of the model is the same as in the compact case: 
the rigid translations. Now however we encounter a technical difficulty: the translation 
group is noncompact and does not preserve the regulator structure. We sidestep this 
problem, introducing the periodic boundary conditions on the faces of the regulator, 
thus restricting to certain compact ``cyclic translation'' group. The new 
``translations'' are obviously not elements of the original translation group, however 
the proposed ``trick'' is well motivated on the heuristic level: as
$(i)$ their action will equal that of the original translations group elements on those 
spin networks which are composed of the set of identical ``copies'' of the portion 
contained within the regulator cell, and $(ii)$ in the regulator removal limit we 
recover the original translation group.

On the technical level the above construction brings us exactly to a compact setting 
$T^3$ topology considered in previous subsections. Thus we proceed with construction 
of the dictionary exactly as before. The results \eqref{eq:p-dict}-\eqref{eq:sigma-inv}
remain true here. In the regulator removal limit the values $p_i$ reach infinity, 
however the plaquets areas are well defined
\begin{equation}\label{eq:sigma-r3}
	\sigma_i = \lim_{n_1,n_2,n_3\to\infty}
		\frac{8\pi\gamma\lPl^2}{n_1 n_2 n_3} \sum_{e\in \{e\}^i} \sqrt{j_e(j_e+1)}
	= 8\pi\gamma\lPl^2 \overline{[ \sqrt{j(j+1)} ]}_{e_i} =: \Delta_i ,
\end{equation}
provided that the limit on the right-hand side exists. it is however a reasonable 
expectation if we consider asymptotically homogeneous state.
As a consequence the relation \eqref{eq:sigma-inv} extends to the noncompact case.

\subsubsection{The isotropic sector}

% rotation invariance
In comparizon to the homogeneous non-isotropic spacetimes considered so far, the 
isotropic ones admit an additional symmetry class (subgroup) -- the rotations.

Implementing these symmetries in the compact $T^3$ case is not possible, 
thus our starting point is the 
noncompact setting of the previous sub-subsection. Here we make one additional 
initial assumption: we restrict the auxiliary metric by requiring that the fiducial
lengths of the edges in all three directions are the same. the flat metric $\oq$ now 
defines the group of rigid rotations. As in the case of the translations, we consider 
them as active diffeomorphisms and average over them the observables used to define
the dictionary.

Due to noncompactness, the only meaningful element of the dictionary is the average 
area of plaquets $\sigma_i$. The rotational transformation of the spin-network can 
be easily parametrized by the Euler angles. As in the case of translations, here 
we can distinguish a discrete group if the rotations by proper angles, which due to 
insensitivity of the area operator to the orientation of the edges can be replaced 
by a group $\Sigma^3$ of permutations of $\ofe_i$. We can then distinguish a quotient 
group $SO(3)/\Sigma^3$.

Let us firsat consider the averageing over (the group of translations and) $\Sigma^3$.
It follows immediately from \eqref{eq:sigma-r3} that 
\begin{equation}\label{eq:sigma-iso1}
	\sigma_{\Sigma} := \sigma_i = \lim_{n_1,n_2,n_3\to\infty}
		\frac{8\pi\gamma\lPl^2}{n_1 n_2 n_3} \sum_{e\in E(\Gamma)} \sqrt{j_e(j_e+1)}
		= 8\pi\gamma\lPl^2 \overline{[ \sqrt{j(j+1)} ]}_{E(\Gamma)} 
		= \frac{1}{3} [ \Delta_1 + \Delta_2 + \Delta_3 ] =: \Delta_{\star} ,
\end{equation}
where this time, we sum over \emph{all} the edges of the spin network. This result 
is identical with the FRW limit of Bianchi I geometry (in LQC) studied in \cite{awe-b1}.

The averaging over (translations and) full $SO(3)$ is slightly more involved. 
\begin{equation}\label{eq:sigma-iso-gen}
	\sigma_R := \sigma_i = \lim_{n_1,n_2,n_3\to\infty}
		\frac{8\pi\gamma\lPl^2}{n_1 n_2 n_3} \int_{SO(3)} \rd\sigma_{SO(3)} 
		\sum_{e\in E(\Gamma)} |\oq_{ab}n^a \ofe(e)_i^b| \sqrt{j_e(j_e+1)} ,
\end{equation}
where $n^a$ is a unit (in $\oq$) vector orthogonal to the plaquet and $\ofe_i^a$ 
is the fiducial triad element tangent to the graph edge $e$. The factor 
$\oq_{ab}n^a \ofe(e)_i^b$ is a consequence of averaging over spatial tranlations as 
only the orthogonal (to $\ofe_i$) ``crossection'' of the plaquet will contribute 
to the average over the translations along $\ofe_i$.
Using the known chart of $SO(3)$ defined by Euler angles (convention 
$Z(\alpha)X(\beta)Z(\gamma)$)\footnote{Instead of rotating the plaquet, we keep 
	it fixed at $z=0$ and rotate the spin network.
} we get
\begin{equation}\label{eq:sigma-iso-gen2}\begin{split}
	\sigma_R &= \lim_{n_1,n_2,n_3\to\infty}
		\frac{8\pi\gamma\lPl^2}{n_1 n_2 n_3} \frac{4}{\pi^2}
		\int_{\alpha\beta,\gamma\in [0,\ldots,\pi/2]}
		\hspace{-1cm}\rd\alpha \sin(\beta)\rd\beta \rd\gamma\,
		\frac{1}{3} [ \sin(\beta)(\cos(\gamma)+\sin(\gamma))+\cos(\beta) ]
		\sum_{e\in E(\Gamma)} \sqrt{j_e(j_e+1)} 
	\\
		&= 4\pi\gamma\lPl^2 \overline{[ \sqrt{j(j+1)} ]}_{E(\Gamma)} 
		= \frac{3}{2}\Delta_{\star} ,
\end{split}\end{equation}
where to write the first equality we split the rotation group onto $\Sigma^3$ and 
the quotient $SO(3)/\Sigma^3$.

As we can see, the contribution from the general angle rotations increases the 
physical plaquet area by a factor $3/2$. This discrepancy is unexpected, since the 
critical energy density (upper bound of the matter energy density operator spectrum)
is a bijective function of the area gap and both Bianchi I and FRW spacetime models 
in LQC provide the same value of that quantity (see \cite{cs-b1} 
versus \cite{acs-aspects}). 
To address this discrepancy we will investigate a bit closer the rotation group used 
in the averaging process: the rotations in fiducial metric $\oq$.

To do so, let us consider a large (classical size) cube of fiducial size $L$. Denote 
its surface area (averaged over the translation group and $\Sigma^3$ to mimic an isotropic
spacetime as close as possible without rotating the graph) in the case 
when its edges are oriented along the triad vectors $\ofe_i$, by $A_{\square}$. 
By repeating the same calculation as in \eqref{eq:sigma-iso-gen2} one can show, 
that upon rotating this cube by Euler angles $(\alpha,\beta,\gamma)$ the surface 
area changes as follows:
\begin{equation}
	A_{\square}(\alpha,\beta,\gamma) 
	= A_{\square} [ \sin(\beta)(\cos(\gamma)+\sin(\gamma))+\cos(\beta) ] .
\end{equation}
This implies in particular, that even if the $j$-label distributions in all the 
directions are the same this surface area \emph{is not invariant under 
the rotations}. The direct consequence of it is the no-go statement: one can not 
build the isotropic spacetime using lattice spin network with the edges oriented 
in directions of one particular vector triad. For that, the large set of spin 
networks oriented in random directions would be needed. Such construction is quite 
easy, if for example one defines the integral Hilbert space structure (discussed in 
sec.~\ref{sec:Hphys}) using the integral measure of $SO(3)$ group. 
In that case however, even the anisotropic universe model would attain the modification
due to integrating over that structure. Indeed, in that case one can rotate the 
original triad $\ofe_i$ and associate with each rotation a lattice state (orthogonal 
with respect to the original lattice state). The set of states supported on new (rotated)
lattices forms now a set of superselection sectors (provided the inner product is 
introduced as in \eqref{eq:Hinteg}). Then
\begin{equation}
	\sigma_i = \int_{SO(3)}\rd\sigma_{SO(3)} \oq_{ab}\ofe_i^a M(g)^b{}_c \ofe_j^c \lim_{n_1,n_2,n_3\to\infty} \frac{8\pi\gamma\lPl^2}{n_1 n_2 n_3}
	\int \rd\sigma(\vec{\epsilon}) \sum_{e\in \{e\}^j_{\Gamma(\vec{\epsilon},g)}}
	\sqrt{j_e(j_e+1)}
\end{equation}
where $g$ is the finite rotation (parametrized by Euler angles) and 
$M(g)$ is the rotation matrix corresponding to it. The symbol $\Gamma(\vec{\epsilon},g)$
denotes here graphs oriented along the fiducial triad $\ofe_i$ rotated by $g$
(that is a support of a superselection sector of lattices oriented in the rotated triad) 
and belonging to the superselection sector labeled by $\vec{\epsilon}$. 
Using again the $SO(3)$ chart defined by Euler angles gives then
\begin{equation}
  \sigma_i = \frac{\gamma\lPl^2}{\pi} 
	\int_{0}^{2\pi} \rd\alpha \int_{0}^{\pi} \sin(\beta)\rd\beta \int_{0}^{2\pi} \rd\gamma\, 
		\sum_{k=1}^{3} (e_i \cdot M(\alpha,\beta,\gamma) e_k)
  \overline{[ \sqrt{j(j+1)} ]}_{e_k,\vec{\epsilon}}(\alpha,\beta,\gamma) =: \tilde{\Delta}_i,
\end{equation}
where the rotation matrix $M$ are now expressed in terms of Euler angles and the 
averages over the translation superselection sectors of $j$-labels (originally 
defined in sec.~\ref{sec:int}) are now defined separately for each rotation 
superselection sector labeled by $(\alpha,\beta,\gamma)$. 

In the case, the averages of $j$-labels of edges in the same direction over the 
graph (within a single superselection sector) are equal, a simple calculation shows that 
\begin{equation}
	\sigma_i = 4\pi\gamma\lPl^2 \sum_{i=1}^3 \overline{[ \sqrt{j(j+1)} ]}_{e_i,\vec{\epsilon}}
	= \frac{1}{2} \sum_{i=1}^3 \bar{\Delta}_i  ,
\end{equation}
where $\bar{\Delta}_i$ is defined in \eqref{eq:avg-sigma}.
This leads exactly to the correction of the Bianchi I plaquet area by a multiplicative 
factor $3/2$
restoring the consistency with the isotropic limit.\footnote{In this case the averaging 
	over rotations can be performed as averaging of $j$-labels between the rotation 
	superselection sectors.
}

\subsection{The physical consequences}

For both construction of separable physical Hilbert space (the single superselection 
sector and the integral one) we reached the same conclusion. Given a ``lattice'' LQG 
state specified in sections~\ref{sec:lat} or~\ref{sec:avg} (as well as its 
extension discussed in previous subsection) the Bianchi I LQC state 
mimicking it has to have the single plaquet area is proportional to the average of 
$j$-labels (in appropriate direction) of the original LQG state. The time dependence 
of the latter depends in turn on $(i)$ the choice of the (initial) LQG state and 
$(ii)$ on the statistics of the particular Hamiltonian used to generate the time 
evolution. In particular, any model following from strictly graph preserving Hamiltonian
will have 
\begin{equation}
	\sigma_i \propto p_i
\end{equation}
which will lead to original Bojowald's prescription in LQC \cite{b-homo1,*b-homo2,*bdv-homo}. 
On the other hand, the studies of the noncompact model, show that in low energy limit 
$\sigma_i$ should be constant. As a consequence, for the class of states considered 
in this article the average $j$-label, more precisely $\Delta_j$ should approach constant
in low curvature limit, thus the expansion of the spacetime in the process of dynamical 
evolution should follow from increasing the number of spin-network nodes rather 
than the $j$-labels. 

This conclusion is supported by our intuitive understanding to the 
physical distance. Essentially, by examining the definitions of meter and second one sees,
that the definition of a distance unit can be recast as a certain number of the spatial 
oscillations of the electromagnetic field corresponding to the photon of certain energy 
(defined in turn by particular spontaneous emission process). On the other hand, 
the coupling of matter to gravity in LQG leads to theory, where the matter degrees of freedom
are represented by quantum labels living on the nodes (vertices) or the edges of the 
spin network (depending on the type of matter). This leads to an intuition, that the 
physical distance should be proportional to the ``number of inhomogeneities'' 
given interval is able to accommodate, thus should be proportional to the number $n_i$ 
of the spin-network edges. 

The observations from the noncompact extensions and the presented intuitions imply, 
that from physically viable models mimicking the cosmological spacetime by LQG 
semiclassical state one should expect the average $j$-label to remain constant at low 
curvatures at least in the leading order. This consistency however does not fix the 
asymptotic value of $\Delta_i$, which however is still subject to the constraints 
following from observations in high energy particle physics, although the upper bounds 
following from it are too huge to be useful.
At present it seems that the more precise values can be provided only by statistical 
analysis of the dynamical evolution within specific frameworks of (or approximations to)
LQG.

One of the ways of determining that value from genuine LQG is provided by the interface 
of Chern-Simons theory with LQG used to evaluate black hole entropy \cite{abck-entr,%
*abk-entr,dl-entr,cdpfb-entr,*cdpfb-entr-det,abfbdpv-prl,bv-entr,epn-entr,bl-entr}.
Indeed, the comparizon of two statistical calculations in \cite{abck-entr,*abk-entr} and 
\cite{dl-entr} shows, that the edges with $j>1/2$ provide significant contribution to 
the surface areas, thus the average ``area gap'' $\Delta_i$ should be detectably 
larger than $\Delta$. On the other hand, more detailed combinatorial analysis of 
the relevant statistics \cite{abfbdpv-count} shows that for small areas the BH 
entropy features a ``stair-like'' structure which on heuristic level can be 
interpreted as the existence of ``quantum of the area''. The numerical simulations 
\cite{cdpfb-entr,*cdpfb-entr-det} determined it to approximately equal 
$\Delta' \approx 7.565\lPl^2$, which would give the average 
$\overline{\sqrt{j(j+1)}} \approx 1.267$ (roughly corresponding to average 
$\overline{j}\approx 0.86$). The stair-like entropy structure dissipates
for larger area due to dispersion in $j$ and the complicated nature of the spectrum of
area operator \cite{bv-degen}, however these are the low areas (small number of 
graph edges intersecting 
the area) where any effect of $j$ distribution, especially any ``peakness'' of it, 
should be the most easy to observe.

While heuristic, the above argument provides a strong indication that, while the LQC 
area gap will not correspond exactly to $\Delta$, it will remain of the same order.
Given an interface constructed here it further suggests a specific correction 
to that area gap\footnote{The idea that the value $\Delta'$ should replace
	the LQC area gap has been originally suggested \cite{dpfb-oral} by authors of 
	\cite{cdpfb-entr}. It was however subsequently abandoned due to lack (at that time) 
	of justification for such choice.
}.
Its value can be determined more precisely via use of the same statistical methods 
originally applied to find the distribution of the magnetic spins 
\cite{abck-entr,*abk-entr,dl-entr,m-entr}. This is however beyond the scope of this article.

\section{Conclusions}

We have considered the relation between the loop quantum cosmology physical states
and the physical states of full loop quantum gravity possibly representing homogeneous
universe on the kinematical level (that is without controlling the consistency of 
dynamical predictions of LQG and LQC frameworks). The LQG has been applied at 
its genuine level, without any simplifications.
Using the observable quantities well defined both in loop quantum gravity and 
cosmology we constructed a precise interface between these two frameworks for the 
class of Bianchi I spacetimes of toroidal spatial topology. To define this interface 
we used the specific subspace of genuine LQG states, distinguished by minimal 
selection criteria allowing to define the necessary components of the LQC auxiliary 
structure for these states and motivated by construction originally proposed in 
\cite{awe-b1}

In the case the physical state is supported on one spin network we assumed that 
the spin networks supporting the state are topologically equivalent to a sub-networks 
of the regular lattices. When the state is composed out of a continuum of states 
on distinct spin networks we further provided a notion of congruence of the embedding 
manifold by parallel lattices. This criterion, being the sole restrictive condition on 
the LQG physical Hilbert space allowed to provide a precise notion of global 
orthogonal\footnote{The meaning of orthogonality has been defined subsequently by 
	no longer restrictive partial gauge fixing.
} directions on the spatial manifold. 

This structure has proven to be sufficient to define all the remaining auxiliary 
LQC structure necessary to construct a dictionary. The embedding manifold has been 
equipped with LQC fiducial background metric via partial gauge fixing. Upon this fixing 
the spin network supporting the state become regular lattice. The choice of (metric) 
regularity of the lattice was however not relevant in further construction of the 
dictionary, being instead a matter of convenience.
Indeed, it was shown, that averaging over the diffeomorphisms preserving the global 
orthogonal directions leads to the same results.

Given the selected class of spin network and the auxiliary structure provided by 
the partial gauge fixing the precise dictionary was constructed, where to identify 
the states in two frameworks we used two classes of observables: areas of global 
$T^2$ slices of the spatial (embedding) manifold -- the maximal surfaces and the 
areas of the square plaquets defined by the minimal loops of the LQG spin network.
This was achieved by averaging the LQG area operators corresponding to these surfaces
over the diffeomorphism transformations remaining after the gauge fixing 
--the rigid translations-- which were considered as active transformations.
In the process no restrictions regarding the distribution of the quantum numbers 
($j$-labels) on the spin network graph have been made.

The result of this identification associated the LQC area of each plaquet with the 
average of $j$-labels (orthogonal to the plaquet) of the LQG spin network. No 
restrictions stronger that this relation \eqref{eq:sigma1}, \eqref{eq:sigma-inv} 
have been found. 
From there we concluded, that the particular association of the values of these areas 
as the LQC phase space functions (known as LQC prescription choice) depend solely 
on the statistical properties of the Hamiltonian (constraint) generating the evolution 
of the spin network and are not restricted by the principles imposed in construction 
the the LQG-LQC interface. In particular both the Bojowald's $\mu_o$ prescriptions 
and the so called \emph{improved dynamics} can be a priori realized.

The results were further extended to the case of flat Bianchi I universe of topologically
$\re^3$ spatial slices, where the relation found in $T^3$ case persisted unmodified 
\eqref{eq:sigma-r3}. this relation and the consistency requirements on the LQC framework
(existence of well defined infrared regulator removal limit) implies some restrictions
on $j$-label statistics of LQG: for the class of states used to build the interface 
the averages of $j$-labels have to remain constant on low energy (gravitational field) 
limit.

The studies were further extended to the case of isotropic flat FRW universe, where the 
plaquet area operators were further averaged over additional symmetries admitted by 
these classes of spacetimes -- the rotations. This process related the (now unique) 
plaquet area with the average of $j$-labels over \emph{all} the edges of the spin network
graph \eqref{eq:sigma-iso1}. Two levels of implementation of the symmetries were considered:
averaging over a discrete group of rotations by proper angles and the full $SO(3)$.
Studies of the former case lead to the FRW limit of Bianchi I cosmology consistent with 
the analogous limit found in \cite{awe-b1}. In the latter it was found, that due 
to contributions of all the graph $j$-labels in the case the rotation angles differ 
from (multiples of) proper ones the area of the plaquet is larger by a factor $3/2$.

Due to apparent discrepancy of the above result 
with the FRW limit of Bianchi I found in \cite{awe-b1} 
the changes of areas in considered model have been investigated up close. Consequently, 
it was found, that the set of lattices oriented in directions of one distinguished 
triad is insufficient to support the states accurately reproducing an isotropic 
spacetime. Consequently, an extension of the Hilbert space using the integral 
structure defined by the group of rotations was proposed. It was further shown 
that on the extended space the discrepancy is cured and the single plaquet area 
in the models of Bianchi I universes is also increased by a factor $3/2$.

The found results have been finally confronted with the heuristic estimates of 
the $j$-statistics following from studies of black hole entropy in LQG. The 
dictionary constructed in this article indicates that associating the plaquet areas 
with the minimal nonzero LQG area is accurate only in cases when the spin network statistics 
makes $j>1/2$ non-generic (zero measure contribution). On the other hand, the 
(known in literature) heuristic results following from numerical analysis of 
black hole entropy provide a natural (from the point of view of the constructed 
dictionary) estimate on the $j$-label averages. This estimate leads again to the 
\emph{constant area gap} principle of improved dynamics. it indicates however a slightly 
different value of this area gap, corresponding to the LQC critical energy density
$\rho_c\approx 0.19\rho_{\Pl}$. This value, while lower than the original LQC critical 
energy density ($\approx 0.41\rho_{\Pl}$) remains at the same level of magnitude.

The general purpose for constructing the above dictionary is to provide a viable 
tool of analyzing the cosmological limit of more advanced models aimed towards 
controlling or approximating the LQG dynamics (see of example \cite{we-lat,ac-red,*ac-let}. Since
$(i)$ the (restrictive) selection criteria are precisely controlled here and the formalism
remains relatively general, and $(ii)$ the formalism is adaptable to majority of prescriptions 
in defining the Hamiltonian (constraint) in LQG, it can be applied to a wide variety of models. 
It allows to extract the cosmological degrees of freedom out of such models 
in a precise way, further providing a tool for validating the initial assumptions 
selected in their construction (like for example the statistical averages 
of $j$ distributions).
Through the consistency conditions on LQC in case of noncompact universes it 
also provides a tool for consistency control of the LQG models.

At this point it is necessary to remember, that the interface relies on quite 
strong restriction of the spin-network graph topology. In principle, no such 
restrictions should be made in order for the results to be completely robust. 
Any generalization however, for example using the random graphs \cite{s-rand} 
is extremely difficult, as in such cases the LQC auxiliary structure (being a 
relevant part of the interface) has to emerge on the physical level (via observables)
and may strongly depend on the $j$-label statistics of the physical states, which
in turn is decided by the details in Hamiltonian (constraint) construction.

\begin{acknowledgments}
	Author thanks Edward Wilson-Ewing for discussions and helpful comments and 
	especially to Emanuele Alesci for extensive discussions and the encouragement to 
	write this article.
	This work has been supported in parts by the Chilean FONDECYT organization under 
	regular project 1140335 and the National Center for Science (NCN) of Poland research 
	grant 2012/05/E/ST2/03308 as well as by UNAB via internal project DI-562-14/R.
\end{acknowledgments}

\appendix

\section{\label{app:stat}Averaging over diffeomorphisms}

Consider a $1$-dimensional lattice ($n$ edges) spanned across the interval $[0,1]$ 
with uniform random distribution of the vertices and each edge equipped with value $x_j$. 
Consider further the ``average'' of some function $f(x_i)$ weighted by the length 
of each edge
\begin{equation}
	\overline{f} = \sum_{i=1}^n f(x_i) l_i
\end{equation}
The probabilistic space of the edge length distribution is the $n$ dimensional 
romboid
\begin{equation}
	\sum_{i=1}^n l_i = 1 ,\qquad \forall i\in\{1,\ldots,n\}:\ l_i>0 ,
\end{equation}
with the measure $\rd\sigma = \rd l_1 \ldots \rd l_n$. The volume of this romboid is
$V_n = 1/n!$. The average value $\langle l_i\rangle$ of $l_i$ is the ratio of the 
volume of the romboid over the $n-1$ dimensional volume ot its base, which is $V_{n-1}$. 
As a consequence we have
\begin{equation}
	\langle\overline{f}\rangle = \sum_{i=1}^n f(x_i) \langle l_i\rangle 
	= \frac{1}{n}\sum_{i=1}^n f(x_i) ,
\end{equation}
which corresponds precisely to the case, where the vertices of the lattice are 
distributed uniformly.

\bibliographystyle{JHEP}
\bibliography{p-embedding}

\end{document}